\documentclass[a4paper, amsfonts, amssymb, amsmath, reprint,
showkeys]{revtex4-1}
\usepackage{graphicx}
\usepackage{epsfig}
\usepackage{epsf}
\usepackage[english]{babel}
\usepackage[utf8]{inputenc}
\usepackage[colorinlistoftodos, color=green!40,prependcaption]{todonotes}
\usepackage{mathtools}
\usepackage{amsthm}
\usepackage{mathtools}
\usepackage{physics}
\usepackage{xcolor}
\usepackage{graphicx}
\usepackage[left=23mm,right=13mm,top=35mm,columnsep=15pt]{geometry} 
\usepackage{adjustbox}
\usepackage{placeins}
\usepackage[T1]{fontenc}
\usepackage{lipsum}
\usepackage{csquotes}
\usepackage[pdftex, pdftitle={Article}, pdfauthor={Author},colorlinks,citecolor=blue,linkcolor=blue,urlcolor=blue]{hyperref}
\def\no{\notag}
\newlength{\figwidth}
\newcommand{\fg}[3]
{\begin{figure}[tb]\vspace*{-0cm}\centerline{\includegraphics[width=\figwidth]{#1}}\vskip
-0.2cm \caption{#3}\label{#2}\end{figure}}
\newcommand{\fgs}[3]
{\begin{figure*}[tb]\vspace*{-0cm}\centerline{\includegraphics[width=0.7\textwidth]{#1}}\vskip
-0.2cm \caption{#3}\label{#2}\end{figure*}}

\newcommand{\bk}{\mathbf{k}}
\newcommand{\br}{\mathbf{r}}

\newcommand{\vl}{\vec{\lambda}}
\newcommand{\vL}{\vec{\Lambda}}
\newcommand{\si}{\sigma}
\def\vA{\mathbf{A}}
\def\cA{\mathcal{A}}

\def\phdag{{\phantom \dagger}}
\def\Jt{\tilde{J}_H}
\newcommand{\ltappr}{{{\lower2pt\hbox{$<$} } \atop \widetilde{ \ \ \ }}}
\newcommand{\gtappr}{{{\lower4pt\hbox{$>$} } \atop \widetilde{ \ \ \ }}}
\newcommand{\dg}{^{\dagger}}

\newcommand{\hmat}[1]{\left(\begin{matrix}  #1 \end{matrix}\right)}

\begin{document}
\title{{Luttinger sum rules and spin fractionalization in the SU($N$)
Kondo lattice}}

\author{Tamaghna Hazra}
\affiliation{Center for Materials Theory, Rutgers University, Piscataway, New Jersey, 08854, USA}
\author{Piers Coleman}
\affiliation{Center for Materials Theory, Rutgers University, Piscataway, New Jersey, 08854, USA}
\affiliation{Department of Physics, Royal Holloway, University of
London, Egham, Surrey TW20 0EX, UK}

\begin{abstract}
We show how Oshikawa’s theorem for the Fermi surface volume of the
Kondo lattice can be extended to the SU$(N)$ symmetric case. By
extending the theorem, we are able to show that the mechanism of Fermi
surface expansion seen in the large $N$ mean-field theory is directly
linked to the expansion of the Fermi surface in a spin-$1/2$ Kondo
lattice. This linkage enables us to interpret the expansion of the
Fermi surface in a Kondo lattice as a fractionalization of the local
moments into heavy electrons. Our method allows extension to a pure
U(1) spin liquid, where we find the
volume of the spinon Fermi surface by applying a spin-twist, analogous
to Oshikawa's flux insertion. Lastly, we discuss the possibility of
interpreting the FL$^*$ phase characterized by a small Fermi surface
in the absence of symmetry breaking, as a non-topological coexistence of
such a U(1) spin liquid and an electronic Fermi liquid.
\end{abstract}

\maketitle

\section{Introduction}

Two decades ago,
Oshikawa\cite{oshikawa2000} applied the
Lieb-Schultz-Mattis approach\cite{lieb1961}
to the Kondo lattice, using its
response to a flux insertion to demonstrate that its
Fermi surface volume counts
the combined density of electrons and
local moments. Although the
expansion of the Fermi surface in the Kondo lattice had been
informally
established from 
arguments of continuity based on the Anderson lattice model\cite{Martin82},
from the large $N$ limit of the Kondo
lattice\cite{read1983,auerbach,indranil05,coleman2016},
Oshikawa's result provided a rigorous foundation for the Fermi surface
expansion in a strict, S=1/2 system Kondo lattice.

Curiously, in the twenty years that have elapsed since this hallmark development, Oshikawa's result has not been generalized to higher group symmetries. Here we show that this generalization is readily
established for a family of SU$(N)$ Kondo lattices. The key result,
is that for local moments
in an antisymmetric representation of the
group constructed from $Q$ elementary spinons, a Fermi liquid
ground-state will have an expanded
Fermi surface volume $V_{FS}$
given by
\begin{equation}\label{voila}
Nv_{{\rm c}}\frac{V_{FS}}{(2\pi)^{D}} = {n_{e}+ N_{S}{Q}},
\end{equation}
where $n_{e}$ and $N_{S}$
are respectively,  the number of electrons and number of local moments
per unit cell of volume  $v_{\rm c}$.
For all $N$, the electronic Fermi surface
expands to incorporate the number of
elementary spinons forming the local moments, and
by increasing $N$ to arbitrarily large values, we can link
Oshikawa's original result to the basin of attraction of
large $N$ field theoretic approaches to the Kondo lattice\cite{read1983,auerbach,coleman2016}.
The importance of this link, is that the Kondo fractionalization
of local moments into {\sl charged } heavy fermions, inferred field theoretically,
is rigorously confirmed.

One of the unexpected outcomes of our analysis, is the discovery that
Oshikawa's flux attachment method can also be applied to
spin liquids~\cite{anderson1987,affleck1988}.  Previously, it was assumed that since spin liquids are neutral,
they are immune to flux attachment, stimulating
an alternative topological interpretation of spin-liquid ground-states
in co-existence with a Fermi liquid.
However, because the unitary transformation that
attaches a flux involves both a charge and a
spin-twist of the wavefunction, a spin-liquid is sensitive to the
flux attachment.  This enables us to show that a U(1)
spin liquid in an SU$(N)$ Heisenberg model,
will have a Fermi surface volume determined by purely by
the number of spinons in the representation, i.e
\begin{equation}\label{voila2}
Nv_{{\rm c}}\frac{V_{FS}}{(2\pi)^{D}} = N_{S}{Q},
\end{equation}
This result suggests that fractionalization in a U(1) spin liquid and
the Kondo lattice does not require a topological interpretation, i.e
that fractionalization and topology are not inevitably tied
together.

The outline of this paper is as follows. In Section~\ref{secDeriv}, we
derive the Luttinger sum rule for the SU$(N)$ Kondo Lattice.  In
Section~\ref{secFrac}, we interpret the result as a signature of
spin fractionalization, cementing an
intuition derived from the large-$N$ mean-field theories as a general
feature of the Kondo lattice. In Section~\ref{secKondoH} we show how the method can be extended to a
Kondo Heisenberg model.
In Section~\ref{secSpinLiq}, we discuss the role of
spin-exchange interactions and identify the spinon Fermi
surface volume of a U(1) spin liquid. Finally in
Section~\ref{secDisc} we discuss whether the co-existence of a spin
and small Fermi surface conduction fluid, to form an FL$^{*}$
requires a topological interpretation.

\fgs{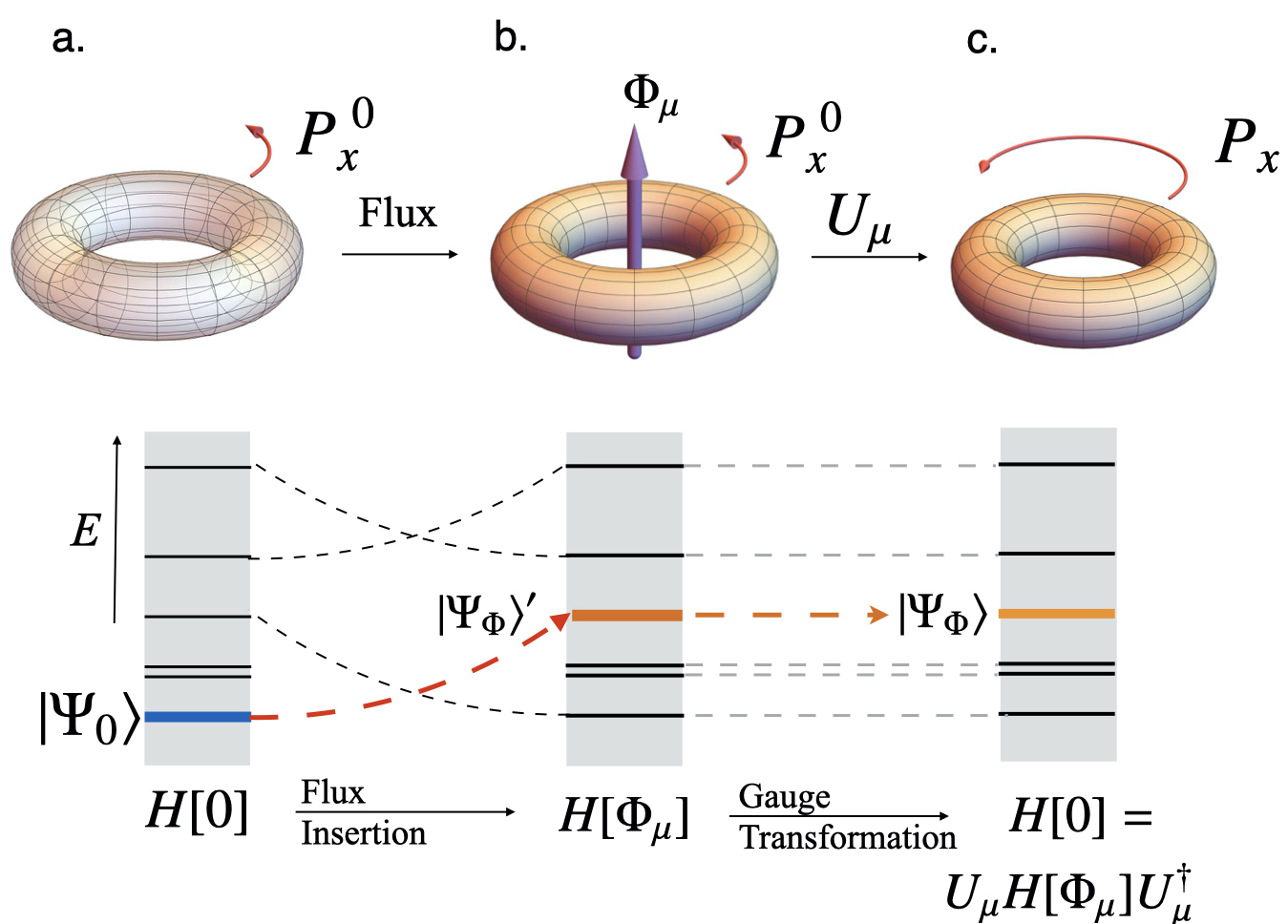}{OshF1}{Flux insertion strategy: a) Initial state $\vert
\Psi_{0}\rangle$ with momentum $P_{x}^{0}$, b) State $\vert
\Psi_{\Phi}\rangle '$  after flux insertion for electrons with spin
component $\mu$,  has unchanged canonical
momentum, c) after gauge transformation, $\vert \Psi_{\Phi }\rangle =
U_{\mu}\vert \Psi_{\Phi }\rangle '$ has canonical momentum $P_{x}$. The
change in momentum $\Delta P_{x}=P_{x}-P_{x}^{0}$ determines the Fermi
surface volume. }

\section{Derivation}\label{secDeriv}
We consider
the SU$(N)$ symmetric  Kondo Lattice
\begin{equation}\label{original}
    H_{KL}=
-\sum_{\br\br'} t_{\br, \br'} c_{\br\si}^\dagger
c_{\br'\si}+
J_K \sum_{\br}   {\vl}_{\br }
\cdot {\vL}_{\br},
\end{equation}
where $c_{\br\si}\dg, (\si=1,N)$ creates an electron at site $\br $,
moving on a $D$-dimensional toroid with
intersite hopping amplitude $t_{\br,\br'}$, with dimensions $L_{x},
L_{y}\dots  L_{D}$.
$ {\vl}_{\br }= c\dg_{\br\si}\vl_{\si\si'}
c_{\br\si' }$ is the electron spin-density at $\br$, where the
 $\vl = (\lambda^{1}, \dots \lambda^{N^{2}-1})$  are the  SU$(N)$
 Gell-Mann matrices. The
$\vL_{\br} = (\Lambda^{1}_{\br},\dots \Lambda_{\br}^{N^{2}-1})$
are the components of the localized
moment at site $\br$.
We shall consider local moments composed of $Q$ elementary spinons, in
an antisymmetric representation of SU(N),
 $\vert \si_{1},\dots
\si_{Q}\rangle = (-1)^{P}\vert \sigma_{P_{1}}\dots
\sigma_{P_{Q}}\rangle $.
The action of
the spin operator $\Lambda^{a}, a= (1,N^{2}-1)$ on these states is then
$ {\Lambda}^a |\sigma_1 ... \sigma_Q\rangle = \sum_{n=1}^Q
|\sigma_1 ... \sigma'_{n}  ... \sigma_Q\rangle \lambda^a_{\sigma'_n\sigma_{n}}.
$

%
%

The SU$(N)$ Kondo lattice has a global U(1)$\times {\rm SU}(N)$ symmetry,
associated with the conserved electron number $N_{e}$ and
magnetization $M^{a}= \sum_{\br} \lambda_{\br}^{a}+
\Lambda_{\br}^{a}$. Of particular interest, are the diagonal components of the
magnetization, $M^{\mu}, (\mu\in [1,N-1])$, which form the Cartan
sub-algebra of the SU$(N)$ group, with  Gell-Mann matrices $\lambda^{\mu}_{\sigma \sigma '}=
(\delta^{\mu\sigma }-1/N)\delta_{\sigma \sigma '}$.

 Oshikawa's strategy (see Fig.~\ref{OshF1})
is to introduce a unit magnetic flux quantum $\Phi_{\mu}=\frac{h}{e}$
 that couples to
the $\mu$th spin component of the Fermi sea, giving rise to a
inductive current which increases
the mechanical momentum by an $\Delta  P_{x}=2\pi/L_{x} \times
V/(2\pi)^D \times V_{FS}^\mu$, directly proportional to the Fermi surface volume.
Since the flux insertion does not change the many-body energy eigenstates,
it is equivalent to a unitary transformation $U_{\mu}$ of the original
Hamiltonian,
$H[\Phi_{\mu}]= U_{\mu}\dg H[0]U_{\mu}$. This enables a direct calculation of the change in the
mechanical momentum due to flux insertion in terms of
microscopic quantities. Equating the direct calculation with the Fermi
liquid result determines the Fermi surface volume.

We now apply this strategy to the SU$(N)$  Kondo lattice.
Flux insertion is achieved
by a Peierls substition
$t_{\br,\br'}\rightarrow
 t_{\br,\br'} e^{-i {\bf
A}^{\sigma }\cdot (\br-\br')},
$ where ${\bf A}^{\sigma }= \delta^{\mu\sigma }\left(\frac{2\pi}{L_{x}} \right)   \hat {\bf x}
$. (Note,  we are using  natural units in which $e=\hbar
=1$ and the dimensions of the unit cell are rescaled to be
unity, so that the unit cell volume $v_{c}=1$.)
This additional gauge field is generated by a large gauge
transformation of the electron fields $U_{\mu}\dg c\dg _{\br \sigma }U_{\mu} = c\dg_{\br\sigma} e^{-i{\bf
A}^{\sigma }\cdot \br }$.  The obvious guess, $ U_{\mu}
    =e^{\frac{2 \pi i}{L_{x}} \sum_{\br} x_{\br}n_{\br
    }^{\mu}}$ does not leave the Kondo interaction
    invariant, but a modified transformation
\begin{align}\label{Umu}
    U_{\mu}
    &=e^{\frac{2 \pi i}{L_{x}} \sum_{\br} x_{\br}\left(n_{\br
    }^{\mu} +\Lambda_{\br}^{\mu}+Q/N\right)},
\end{align}
satisfies this requirement.
This is a generalization of Oshikawa's original
transformation, in which we have replaced the $SU (2)$ generator
$S_{\br}^{z}$
by $\Lambda^{\mu}_{\br}$. We have also added an additional
gauge transformation which multiplies the wavefunction by a factor $e^{\frac{2\pi
i }{L_{x}}x_{\br}Q/N}
$ at each site, which ensures that the unitary transformation preserves
the periodic boundary conditions.
$U_{\mu} (\left\{x_{\br} \right\})=U_{\mu} (\left\{x_{\br} +L_{x}\right\})
$. $U_{\mu}$ is actually a  product of
a U(1) and an SU$(N)$ gauge
transformation: in other words, to selectively impart momentum to the $\mu$th
Fermi surface we must ``twist'' the wavefunction in
charge {\sl and}  spin space.

To see that $U_{\mu}$ commutes with the Kondo interaction we write
$n^{\mu}_{\br}=\lambda^{\mu}_{\br}+
n_{\br}/N$
so that
\begin{equation}\label{}
U_{\mu}
=e^{\frac{2 \pi i}{L_{x}} \sum_{\br} x_{\br}\left[(n_{\br}+Q)/{N}+ M_{\br}^{\mu}
\right]}.
\end{equation}
involves
the electron density $n_{\br}$ and local magnetization $M_{\br}^{\mu}=
\lambda_{\br}^{\mu}+\Lambda_{\br}^{\mu}$,
which both commute with the
Kondo interaction.  To confirm that the transformation also preserves
periodic boundary conditions, we note that if we shift the x-component
of the site at $\br_0$ by $L_{x}$, i.e $x_{\br_0}\rightarrow
x_{\br_0}+L_{x}$,
the unitary transformation picks up an
additional factor $
e^{2\pi i (n^{\mu}_{\br_{0}} +
\Lambda_{\br_{0}}^{\mu}+q)} =
e^{2\pi i (\Lambda_{\br_{0}}^{\mu}+q)}
$, where $q=Q/N$ and we have used the fact that the $n_{\br}^{\mu}$ are integers.
But under a $2\pi$ rotation, an SU$(N)$ spin,
picks up a phase factor
i.e $e^{2\pi i \Lambda_{\br_{0}}^{\mu}} =e^{-2\pi i q}$, so that the
factor $e^{2\pi i (\Lambda_{\br_{0}}^{\mu}+q)} =1$ and the unitary
transformation $U_{\mu}$ preserves periodic boundary conditions.

\begin{widetext}
Written in full, the Hamiltonian with flux inserted is
\begin{eqnarray}\label{l}
H[\Phi_{\mu}]
 &=&
-\sum_{\br\br'\sigma} t_{\br, \br'}
e^{-i {\bf
A}^{\sigma }\cdot (\br-\br')}
 c_{\br\si}^\dagger
c_{\br'\si}+
J_K \sum_{\br}   {\vl}_{\br }
\cdot {\vL}_{\br},\cr
{\bf A}^{\sigma }&=& \delta^{\sigma\mu}\left(\frac{2\pi}{L_{x}} \right)   \hat {\bf x}.
\end{eqnarray}
The process of flux insertion involves adiabatically increasing $
{\bf A}^{\sigma } (t)=  {\bf A}^{\sigma } e^{-|t| /\tau }
$ from zero at $t=-\infty $ to its full value at $t=0$, taking $\tau
\gg (1/T_{K})$ to be much longer than the inverse Kondo
temperature,
so that the initial eigenstate $\vert \psi^{0}\rangle $ evolves
smoothly into
an excited eigenstate $\vert \psi_{\Phi }\rangle '$ of
$H_{KL}[\Phi_{\mu}]$ (see Fig. \ref{OshF1}).

Since translational symmetry is preserved by flux insertion, and since
the exponential
of the canonical momentum $e^{-i P_{x}}$ is the eigenstate of translation, it follows that
the state retains a fixed canonical momentum $P_{x} (t)=P_{x}^{0}$ so
that  under a
translation,
\begin{equation}\label{}
T_{x}|\psi_{\Phi}\rangle' = e^{-iP_{x}^{0}}|\psi_{\Phi
}\rangle'.
\end{equation}
We can obtain the mechanical momentum $P_{x}$ of the final state
$\vert \psi_{\Phi }\rangle '$ by
noting that since this quantity is gauge invariant, it is unchanged
when we  gauge transform back into the original gauge.
Now since
$H_{KL}[0]=U_{\mu}H_{KL}[\Phi_{\mu}]U\dg_{\mu}$,
it follows that
$ \vert \psi_{\Phi}\rangle= U_\mu |\psi_\Phi\rangle'$ is the corresponding
transform of $\vert \psi_{\Phi }\rangle $ back into the original
gauge.
But since the
vector potential is now absent, the mechanical and canonical momentum coincide
and can be determined from a translation,
\begin{equation}\label{}
T_{x}\vert
\psi_{\Phi}\rangle = e^{-iP_{x}}\vert \psi_{\Phi}\rangle.
\end{equation}
Since $T_{x}\vert \psi_{\Phi }\rangle  = (T_{x}
U_{\mu}T_{x}^{-1})T_{x}|\psi_{\Phi }\rangle '$, it follows that
\begin{equation}\label{eqmomentum}
 e^{-iP_{x}}\vert \psi_{\Phi}\rangle= (T_{x} U_{\mu}T_{x}^{-1})
e^{-iP_{x}^{0}}|\psi_{\Phi }\rangle '.
\end{equation}
Now $T_{x}U_{\mu}T_{x}^{-1}$ describes the effect of translating
the operator $U_{\mu}$ by one lattice spacing in the $\hat x$
direction, so that
\begin{eqnarray}\label{trans}
(T_{x} U_{\mu}T_{x}^{-1})
=
\exp\left[\frac{2 \pi i}{L_{x}}
\sum_{\br} x_{\br}\left(n_{\br+\hat{x}}^{\mu} +\Lambda_{\br+\hat{x}}^{\mu}+q\right)\right]=\exp\left[\frac{2 \pi i}{L_{x}} \sum_{\br} x_{\br-\hat {x}}
\left(
n_{\br}^{\mu}+\Lambda_{\br}^{\mu}+q\right)\right],
\end{eqnarray}
where inside the sum, we have shifted the x-coordinate of the position vectors $\br$,
$\br\rightarrow \br-\hat x $. Now naively we might expect $x_{\br-\hat x}=x_{\br}-1$.
However this is not the case with sites on the first layer of the crystal,
for
in this case $x_{1-1}= x_{0}$, but the periodic boundary conditions
mean that $x_{0}=x_{L_{x}}= L_{x}= x_{1}-1+L_{x}$. Thus
in general,  $x_{\br-\hat  x}= x_{\br}-1+ L_{x}\delta_{r_{1},1}$.
Substituting this into (\ref{trans})
we obtain
\begin{eqnarray}\label{l}
(T_{x} U_{\mu}T_{x}^{-1})
&=&\exp\left[\frac{2 \pi i}{L_{x}} \sum_{\br}\left(x_{\br}-1+
    L_{x}\delta_{r_1,1}\right)\left(n_{\br}^{\mu}+\Lambda_{\br}^{\mu}+q\right)\right]\cr
&=& \exp\left[\frac{2 \pi i}{L_{x}} \sum_{\br}\left(
    L_{x}\delta_{r_1,1}-1
\right)\left(n_{\br}^{\mu}+\Lambda_{\br}^{\mu}+q\right)\right]U_{\mu}
\cr
&=& \exp \left[2 \pi i \sum_{\br_{\perp} }
(n_{1,\br_\perp}^{\mu}+\Lambda_{1,\br_\perp}^{\mu}+q)
\right]
\exp\left[-\frac{2 \pi i}{L_{x}} \sum_{\br}
\left(n_{\br}^{\mu}+\Lambda_{\br}^{\mu}+q\right)\right]U_{\mu}
\end{eqnarray}
The first term derives from the crystal boundary at $x_{\br}=1$,
derived from the shift of the x-coordinates by $L_{x}$.
However, since we have chosen a gauge where $U_{\mu}$ is invariant
under such co-ordinate shifts, this pre-factor is unity ($e^{2\pi i
 (n_{1,\br_\perp}^{\mu}+\Lambda_{1,\br_\perp}^{\mu}+q)}
=1$). Our final answer for the translated $U_{\mu}$ is then
\begin{equation}\label{key_identity}
(T_{x} U_{\mu}T_{x}^{-1})
= \exp\left[-\frac{2 \pi i}{L_{x}} \sum_{\br}
\left(n_{\br}^{\mu}+\Lambda_{\br}^{\mu}+q\right)\right]U_{\mu}.
\end{equation}
We note that this answer is also obtained with Oshikawa's original
choice of $U_{\mu }= e^{\frac{2\pi i }{L_{x}}\sum_{\br}x_{\br}
(n_{\br}^{\mu}+\Lambda_{\br}^{\mu})}$, but in this case, the q-dependence
derives from the boundary term.
\end{widetext}

From \eqref{eqmomentum} it then follows that
\begin{equation}\label{eq5}
 e^{-iP_{x}}\vert\psi_{\Phi}\rangle = e^{-iP_{x}^{0}}
 e^{-\frac{2 \pi i}{L_{x}} \sum_{\br}\left(n_{\br}^{\mu}+\Lambda_{\br}^{\mu}+q\right)}
\vert\psi_{\Phi}\rangle,
\end{equation}
i.e, flux insertion changes the mechanical
momentum  by
\begin{align}
    \Delta P_{x}=
\frac{2 \pi}{L_{x}} \sum_r\left(n_{\br}^{ \mu}+\Lambda_{\br}^{
\mu}+q\right).
\end{align}
For $N_s$ spins per unit cell,
\begin{align} \label{eqdelP1}
    \Delta P_{x}=\frac{2 \pi}{L_{x}} V
    \left[\nu^{\mu}+N_{s}\left(m^{\mu}+q\right)\right] \quad \text { mod
    } 2\pi,
\end{align}
where $V=L_x L_y \ldots L_D$ is the system volume,
while $\nu^{\mu}=(1/V)\sum_\br n_{\br}^{\mu}$
and $m^{\mu}=(1/V) \sum_{\br}\Lambda_{\br}^{\mu}$ are the  $\mu$-th
filling fraction and magnetization respectively.

Alternatively, if we assume a Fermi liquid ground state, we can
compute the change in momentum by observing
that coupling to the gauge potential shifts the momentum of each $\mu$-quasiparticle by $2\pi/L_x$, so that $\Delta P_{x}=\frac{2 \pi}{L_{x}} N_{F}^{\mu} $
where $N_{F}^{\mu}$ is the number of $\mu$-quasiparticles. The quasiparticle number operator $\tilde{n}_{\bk \mu}$ is conserved in a Fermi liquid and jumps from 1 to 0 across the Fermi surface. This allows us to relate the shift in momentum to the volume of the $\mu$-Fermi surface $V_{FS}^{\mu}=N_{F}^{\mu} (2\pi)^D/V$
\begin{align} \label{eqdelP2}
    \Delta P_{x}=\frac{2 \pi}{L_{x}} V \left[ \frac{V_{FS}^{\mu}}{(2\pi)^D} \right].
\end{align}
Comparing Eq.~\eqref{eqdelP1} and Eq.~\eqref{eqdelP2} we find
\begin{equation}\label{}
V \frac{V_{FS}^{\mu}}{(2\pi)^D} = V (\nu^\mu+N_s(m^\mu + q)) + n_x
L_x,
\end{equation}
with $n_x\in \mathbb{Z}$. Now since the remainder term $n_{x}L_{x}$
can be calculated for a flux threading in any
of the $D$ directions, the remainder is also equal to 
$n_{y}L_{y}, \dots n_{D}L_{D}$, where the $n_{i}$ ($i=1,D$) are distinct
integers for each direction. But since integer remainder is
independent of direction, $n_{x}L_{x}= n_{y}L_{y}=\dots
n_{D}L_{D}$. If we choose the 
$L_x, L_{y}\dots  L_{D}$ to be coprime (no common denominators),
it follows that $n_{x}$ is proportional to each of the
$n_{y}L_{y}, \dots n_{D}L_{D} $, so that the 
it follows that
the remainder is a multiple of the full product, i.e the volume  $V=L_{x}\dots
L_{D}$. Factoring out the volume $V$, we obtain
\begin{align}    \label{eqfinal}
    \frac{V_{FS}^{\mu}}{(2\pi)^D} =\nu^\mu+N_s(m^\mu + q)+n
\end{align}
Since the Fermi surface volume is an intensive quantity, the remainder $n$ is independent
of the convenient choice of mutually coprime boundary lengths,
and  Eq.~\eqref{eqfinal} is valid in the thermodynamic limit.

Finally, if we trace over all $N$ Fermi surfaces, since the members of
the Cartan sub-algebra are traceless, it follows that
$\sum_{\mu}m^{\mu}=0$ so that
\begin{eqnarray}\label{l}
  N v_{c}
\frac{V_{FS}}{(2\pi)^D} =n_{e}+N_s Q\label{eqfinal}
\end{eqnarray}
where $n_{e}= \sum_{\mu}\nu_{\mu}$ and we have restored the
engineering dimensions of the unit cell volume $v_{c}$, and have
dropped the integer remainder $p=nN$, with the understanding that the
Fermi surface volume is only defined mod $(2\pi)^{D}$.

\section{The link with fractionalization}\label{secFrac}

Traditionally, the localized spins of a Kondo lattice are written in
terms of an Abrikosov pseudo-fermion representation
\begin{align}
    \Lambda_\br^a = f_{\br\si}^\dagger \lambda^a_{\si\si'}
    f_{\br\si'}^\phdag, \qquad (a=1,N^{2}-1)
\end{align}
with a constraint on the local $f$-fermion (spinon) density
$n_{\br}^{(f)}=\sum_\mu f_{\br\mu}^\dagger f_{\br\mu}^\phdag
=Q$ which determines the number of spinons contained in the
$Q$th antisymmetric representation of SU(N).
With hindsight, we now see that since the
constraint commutes with evey operator involved in the proof, we could
have used this representation from the outset, but by
tacitly avoiding doing so, we avoided any lingering
concerns about the constraint.

In the Abrikosov representation, the
Kondo Lattice Hamiltonian takes the form~\cite{read1983}
\begin{align}
	H_{KL}=-\sum_{\br\br'} t_{\br, \br'} c_{\br\si}\dg c_{\br'\si}
-\frac{\tilde{J}_K}{N}\sum_{\br} c_{\br\si}^\dagger f_{\br\si}^\phdag f_{\br\si'}^\dagger c_{\br\si'}^\phdag
\end{align}
which explicitly commutes with the constraint $n_{f\br}=Q$ and the
number of conduction electrons $n_{c\br}$ at site $\br$. With the normalization ${\rm Tr}[\lambda^{a}\lambda^{b}]=
(1-\frac{1}{N})\delta^{ab}$ set by the Cartan sub-algebra,
the coupling constants of the Read-Newns form and the original model~\eqref{original} are related by
$\tilde{J}_{K}= J_{K}(N-1)$.

The Cartan elements are now represented by
$\Lambda^\mu_{\br }=n_{f\br}^{\mu}-Q/N$, so that the gauge
transformation \eqref{Umu} that imposes the flux insertion is given by
\begin{eqnarray}\label{fracky}
U^{\mu}=
\exp \left[\frac{2\pi i}{L_{x}}\sum_{\br} x_{\br}
(n_{c\br}^{\mu }+n_{f\br}^{\mu})
\right].
\end{eqnarray}
\eqref{fracky} is literally, a large gauge
transformation that counts the $f-$spinons as quasiparticles.
The conduction electrons and spinons transform identically under
the flux insertion,
\begin{eqnarray}\label{fluxtransform}
U\dg _{\mu}
\hmat{c\dg _{\br \si}\cr f\dg _{\br\si}}
U_{\mu}
= e^{- i {\bf A}^{\sigma }\cdot \br}
\hmat{c\dg _{\br \si}\cr f\dg _{\br\si}}.
\end{eqnarray}
In other words, the structure of the unitary transformation, forced
upon us by the Kondo coupling, means that
the spinons behave exactly as charged particles under the flux
attachment, consistent with a fractionalization of spins into heavy
electrons in the Fermi liquid phase.
Remarkably then,  the seeds of fractionalization are present in the
original Oshikawa gauge transformation.

The final form of the Luttinger sum rule
\begin{align}
    \frac{V_{FS}^{\mu}}{(2\pi)^D} =\nu^\mu+\nu^\mu_{f}, \qquad {\rm mod} (1)  \label{eqfinalAbrikosovform}
\end{align}
where
$\nu^\mu_{f}=\frac{N_s}{V} \sum_\br n_{f\br}^{\mu}= N_{s} (
m^{\mu}+q)$
is the number of spinons with spin index $\mu$ per unit cell, is not a
surprise, because
the
$U(1)\times SU(N)$ gauge transformation \eqref{fracky} audits every spinon
entangled into the Fermi sea.

Traditionally,
the Kondo Fermi surface expansion is interpreted by
identifying the Kondo Hamiltonian as the
strong coupling renormalization of a periodic Anderson model with
the same filling\cite{Martin82}. 
However, a Kondo lattice Hamiltonian has no knowledge of its high energy
origins.
From a renormalization group perspective,
the Kondo lattice lies on the common
scaling trajectory of many high energy ``microscopic''
Hamiltonians. Indeed, the model is entirely agnostic as to the origin
of the local moments, and they need not have an electronic origin at
all, for instance,
they equally could be nuclear spins, with
a Kondo interaction derived from hyperfine interactions.
The main point is that since the Kondo lattice has no knowledge of its
high energy origins, fractionalization in the Kondo lattice
is an emergent property. This alternate interpretation
allows us to contemplate the possibility that different kinds of spin
fractionalization may develop in the approach to magnetism, or spin
liquid behavior. 

\fgs{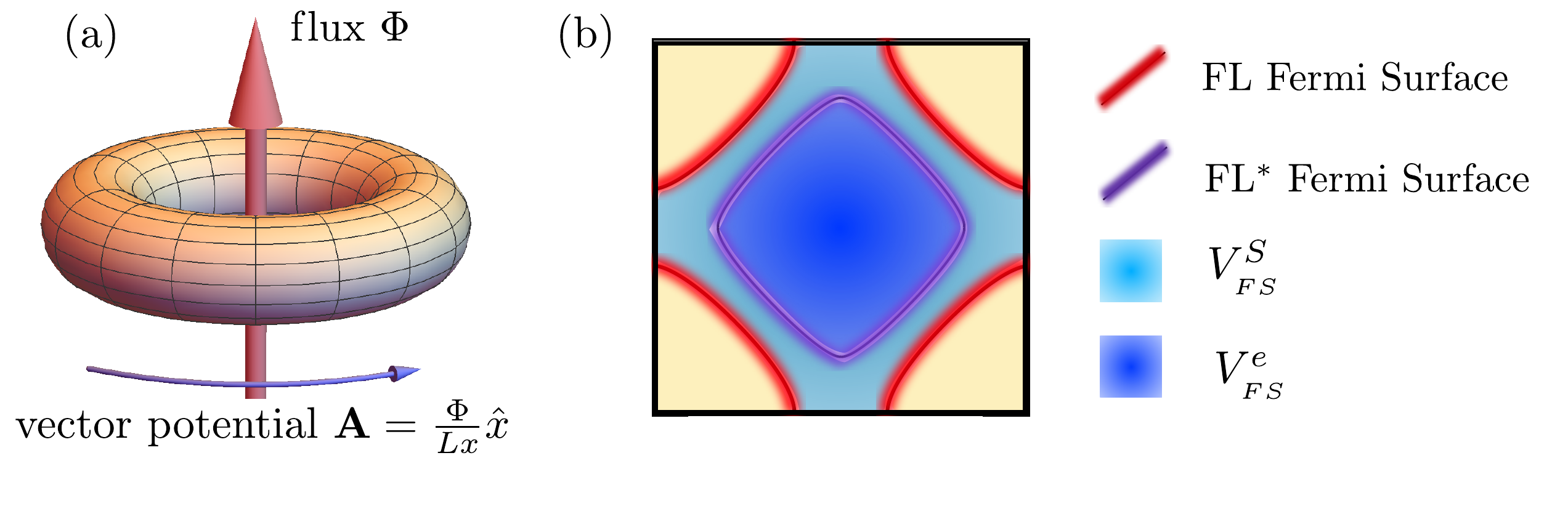}{fSchematic}{ (a) Flux attachment in the Kondo
Heisenberg model.  Threading a flux results in a twist in the U(1)
gauge potential and a twist in the spin orientations, imparting
momentum to the electrons and the spinons. (b) The total momentum is
proportional to the combined Fermi surface volume of the electrons and
spinons (red). In the FL$^*$ phase the spinons decouple from the
electrons to  form a U(1) spin liquid, resulting in a smaller Fermi
surface (purple) that only counts the electrons.}

\section{Kondo Heisenberg Model}\label{secKondoH}

We now consider an extension of our results to a Kondo Heisenberg
model: a Kondo lattice
with additional Heisenberg interactions, $    H_{KH}= H_{KL}+ H_{H}$,
where now
\begin{align}
H_{H}=  \sum_{\langle\br\br'\rangle} J_{\br,\br'}\vec{\Lambda}_\br \cdot \vec{\Lambda}_{\br'}.
\end{align}

From Doniach's original arguments\cite{Doniach}, we know that
for large enough $T_{K}$, the Kondo interaction will
stabilize a Fermi liquid, in which case, we expect Oshikawa's result to
generalize to the Kondo Heisenberg model.
We are particularly interested in the case of frustrated Kondo
lattices, where in the limit of small $T_{K}$, rather than forming a
state of long-range magnetic order, the system develops
into spin liquid, preserving
the Fermi surface of the underlying spinons.  We shall show that
Oshikawa's theorem can be extended to this case.

Naively, one might expect flux insertion
to only affect charge particles, leaving the Heisenberg term
alone. However, the unitary transformation
that accomplishes flux insertion \eqref{Umu},
$U_{\mu}  =e^{\frac{2 \pi i}{L_{x}} \sum_{\br} x_{\br}\left(n_{\br
    }^{\mu}
+\Lambda_{\br}^{\mu}+q\right)}$, adds a charge and a spin-flux to the
system, thus affecting the Heisenberg interaction terms.
Under the gauge transformation, the local
 moments transform under the adjoint representation of SU(N).
To keep  track of these transformations, its simpler to switch
to a Coqblin Schrieffer representation of the local moments,
$\Lambda^{\si\si' }_{\br}
= f\dg_{\br \si }f_{\br \si' }- \frac{Q}{N}\delta_{\si\si'}$,
so that the Heisenberg
interaction takes the form
\begin{equation}\label{}
H_{H} = \frac{1}{N}\sum_{\langle\br\br'\rangle} \tilde{J}_{\br,\br'}
\Lambda^{\si\si'}_{\br }\Lambda^{\si'\si}_{\br' },
\end{equation}
where $\tilde{J}_{\br,\br'}= J_{\br,\br'} (N-1)$.
Under the flux insertion,
$f_{\br \si}\rightarrow
e^{i \vA^\si \cdot\br }
f_{\br\sigma},
$ so that
 under the
gauge transformation \eqref{fluxtransform},
\begin{equation}\label{}
\Lambda^{\si \si' }_{\br}\rightarrow U_{\mu}\dg \Lambda^{\si \si' }_{\br}U_{\mu}=
e^{-i (\vA^\si-\vA^{\si'}) \cdot \br }
\Lambda^{\si \si' }_{\br},
\end{equation}
which describes the transformation of the spin operator under the
adjoint representation of SU(N), corresponding to a
slow twist of the local moments about the ``$\mu$'' axis, created by
the spin component of $U_{\mu}$, through an
angle $2\pi (x/L_{x})$ that increases from $0$ to $2\pi$ across the
sample.

\begin{widetext}
Using these results, we can write
Heisenberg Kondo model with a flux insertion in the $\mu$ spin channel as
\begin{eqnarray}\label{l}
H_{KH}[\Phi_{\mu}]
 &=&
-\sum_{\br\br'} t_{\br, \br'}
e^{-i {\vA^\si}
\cdot (\br-\br')}
 c_{\br\si}^\dagger
c_{\br'\si}+
\frac{\tilde{J}_K }{N}
\sum_{\br}  c\dg_{\br \si  }c_{\br\si'}
\Lambda^{\si \si
}_{\br }
+
\frac{1}{N}\sum_{\langle\br\br'\rangle} \tilde{J}_{\br,\br'}
e^{-i (\vA^\si - \vA^{\si'}) \cdot (\br-\br')}
\Lambda^{\si\si'}_{\br }\Lambda^{\si'\si}_{\br' }.
\end{eqnarray}
The gauge field inside the Heisenberg term
\begin{equation}\label{}
e^{-i \vA \cdot (\br-\br') (\delta_{\si \mu}-\delta_{\si' \mu})}=
e^{-i \vA \cdot (\br-\br') \delta_{\si \mu}}
e^{i \vA \cdot (\br-\br') \delta_{\si' \mu}}, \qquad \qquad (\br' \xrightleftharpoons[\ \si'\ ]{\ \si\ } \br )
\end{equation}
can be interpreted as the product of two Peierls' insertions
associated with a spinon exchange: an
$\si $ spinon moving from $\br'$  to $\br$,
and a $\si' $ spinon moving in the opposite direction.
The derivation and final form of the Luttinger sum rule for the Fermi liquid
now follows precisely the same route as in the Kondo model. In
particular, the key identity \eqref{key_identity} still holds,
allowing us
to generalize the Oshikawa result\eqref{voila} to the
Fermi liquid phases of the SU$(N)$ Kondo Heisenberg model.
\end{widetext}

Since our flux insertion works for arbitrary N, it
allows us to
explicitly examine how the wavefunction $\vert \Psi_{0}\rangle $
evolves at large $N$, allowing us to
the explicit evolution under the flux attachment and subsequent gauge transformation,
\begin{equation}\label{}
\vert \Psi_{0}\rangle {\xrightarrow {\Phi_{0}} }
\vert \Psi_{\Phi}\rangle '{\xrightarrow {U_{\mu}} } \vert  \Psi
\rangle.
\end{equation}
At large $N$, the ground-state wavefunction is accurately determined by
a Gutzwiller wavefunction
\begin{equation}\label{}
\vert \Psi_{0}\rangle  = P_{G} \prod_{\bk \in {\rm FS}, \sigma }
(\alpha_{\bk }c\dg_{\bk \sigma }+ \beta_{\bk }f\dg_{\bk \sigma })\vert  0\rangle \end{equation}
where the product runs over all wavevectors enclosed by the Fermi
surface, and $P_{G} = \prod_{\br}\delta_{n_{f} (\br),Q}$ projects out
the component of the wavefunction with $n_{f} (\br) =Q$ at each site,
while the hybridized operators $\alpha_{\bk \sigma  }c\dg_{\bk \sigma
}+ \beta_{\bk }f\dg_{\bk \sigma }$ define the quasiparticles of the
mean-field Hamiltonian.  In fact, the Gutzwiller projection $P_{G}$ can
be replaced by an average constraint in the large $N$ limit, but here we
shall keep it for greater generality.

In the large $N$ limit, the dynamics of the wavefunction are
determined by  evolution under a time-dependent, translationally
invariant  mean-field Hamiltonian which preserves the
momenta of the quasiparticle states, leaving the Fermi surface
unchanged.
After the flux insertion, the mean-field wavefunction then has the form
\begin{equation}\label{}
\vert \Psi_{\Phi }\rangle'  = P_{G} \prod_{\bk \in {\rm FS}, \sigma }
(\alpha_{\bk \sigma }[\Phi]c\dg_{\bk \sigma }+ \beta_{\bk\sigma
}[\Phi ]f\dg_{\bk \sigma })\vert  0\rangle \end{equation}
where the coefficients $\alpha _{\bk \sigma }[\Phi ]$ and $\beta_{\bk
\sigma }[\Phi]$ differ from their zero field value by terms of order $O
(1/L)$.
Now if we Fourier
transform \eqref{fluxtransform} the transformation of
the electron and spinon fields under $U_{\mu}$, in momentum space is
given by
\begin{eqnarray}\label{fluxtransformk}
U_{\mu}
\hmat{c\dg _{\bk \si}\cr f\dg _{\bk\si}}
U\dg _{\mu}
= \hmat{c\dg _{\bk+ \vA^{\sigma } \si}\cr f\dg _{\bk + \vA^{\sigma
}\si}},
\end{eqnarray}
so that under the unitary transformation $U_{\mu}$, $U_{\mu}\vert
\psi_{\Phi }\rangle ' = \vert \psi_{\Phi }\rangle$ is given by
\begin{equation}\label{}
\vert \psi_{\Phi }\rangle
 =  P_{G}\prod_{\bk \in FS, \sigma }
\left(\alpha_{\bk\sigma  }[\Phi ]c\dg_{\bk + \vA^{\sigma }\sigma }+ \beta_{\bk\sigma  }[\Phi ]
f\dg_{\bk+ \vA^{\sigma }\sigma } \right)\vert  0 \rangle,
\end{equation}
If we translate this state in the $x$ direction, then since
\begin{eqnarray}\label{fluxtransformk}
T_{x}
\hmat{c\dg _{\bk \si}\cr f\dg _{\bk\si}}
T_{x}^{-1}
= e^{-i k_{x}} \hmat{c\dg _{\bk \si}\cr f\dg _{\bk \si}},
\end{eqnarray}
it follows that the momentum of the final state $P_{x}$ is given by
\begin{equation}\label{}
T_{x}\vert \psi_{\Phi }\rangle = e^{-i P_{x}}\vert  \psi_{\Phi
}\rangle,
\end{equation}
where
\begin{equation}\label{}
P_{x }= \sum_{\bk  \in FS, \sigma } (k_{x}+ A_{x}^{\sigma }) =
P_{x}^{(0)}+\frac{2\pi}{L_{x}}\frac{V}{(2\pi)^D} V_{FS}^\mu
\end{equation}
so we see that the shift in momentum per quasiparticle
is precisely $A_{x}^{\sigma }=
\frac{2\pi}{L_{x}}$ in the $\mu$ band.


\section{U(1) Spin Liquid}\label{secSpinLiq}

The fascinating aspect of this result, is that it also allows us to apply the
flux attachment idea to a pure Heisenberg model $H_{H}$. The
Heisenberg model with a spin twist, $H_{H}[\Phi_{\mu}]= U\dg _{\mu}H_{H}U_{\mu}$ is written
\begin{eqnarray}\label{l1}
H_{H}[\Phi_{\mu}]&=&\frac{1}{N}\sum_{\langle\br\br'\rangle} \tilde{J}_{\br,\br'}
e^{-i (\vA^\si-\vA^{\si'}) \cdot (\br-\br') }
\Lambda^{\si\si' }_{\br }\Lambda^{\si'\si  }_{\br' },\cr
\vA^{\sigma }&=& \delta^{\sigma\mu}\frac{2\pi}{L_{x}}\hat x
\end{eqnarray}
and the corresponding gauge transformation is then
\begin{eqnarray}\label{l}
U^{s}_\mu &=& \exp \left[ \frac{2\pi i }{L_{x}}\sum_{\br} x_{\br }
 (\Lambda_{\br}^{\mu}+q)\right]
.
\end{eqnarray}
In this case, the translated gauge transformation takes the form
\begin{equation}\label{}
(T_{x} U^s_{\mu}T_{x}^{-1})
= \exp\left[-\frac{2 \pi i}{L_{x}} \sum_{\br}
\left(\Lambda_{\br}^{\mu}+q\right)\right]U^s_{\mu}.
\end{equation}
so the change in momentum associated with the flux insertion is then
\begin{eqnarray}\label{eqdelPsl}
    \Delta P_{x}&=&
\frac{2 \pi}{L_{x}} \sum_r\left(\Lambda_{\br}^{
\mu}+q\right)\cr
&=&\frac{2 \pi}{L_{x}}  V (m^{\mu}+q)
\quad \text { mod
    } 2\pi,
\end{eqnarray}
where $V= L_{x}L_{y}\dots L_{D}$ is the volume and $m^{\mu}=
\frac{1}{V}\sum_{\br}\Lambda^{\mu}_{\br}$ is the magnetization, and we
have assumed $n_{s}=1$ local moment per unit cell.
Using Abrikosov fermions, $\Lambda_{\br}^{\mu}+q= n_{f\br}^{\mu}$ is
the number of $\mu$- spinons at site $\br$, so we
can interprete $V (m^{\mu}+q) $ as the number of spinon with spin
component $\mu$. In other words, under a flux attachment, each
spinon with spin component $\mu$ in the ground-state acquires a
momentum $\frac{2\pi}{L}$.

A U(1) spin liquid can be thought of as an incompressible neutral Fermi
liquid.  In Appendix~\ref{appMFT}, we demonstrate that such a state is energetically favored in the large $N$ limit over the dimer phase and the $\pi$ flux phase on a square lattice over a range of $q$. To see how its momentum changes under a flux attachment,
consider the  model ground-state
provided by a Gutzwiller wavefunction
\begin{equation}\label{}
\vert \Psi_{0}\rangle  =
P_{G} \prod_{\bk \in {\rm FS}, \sigma }
f\dg_{\bk \sigma }\vert  0 \rangle,
\end{equation}
where, as in the Kondo lattice, $P_{G}=\prod_{\br}\delta_{n_{f}
(\br),Q} $ is a Gutzwiller projection onto states with $Q$ elementary
spinons at each site.
Now the translation operator commutes with
$P_{G}$, and since
$T_{x}f\dg_{\bk \sigma }T_{x}^{-1}= e^{- i k_{x}}f\dg_{\bk
\sigma }$, it follows that this state has the initial momentum
$P_{x}^{(0)} = \sum_{\bk \in {\rm FS}, \si} k_x$.
In the large $N$ limit (see Appendix~\ref{appFlux}),
the time-evolution of the state is given by a time-dependent
mean-field Hamiltonian that is explicitly translationally invariant,
so that under  a flux attachment, the 
canonical momenta of the spinons are entirely unchanged.
In a one-band fluid of spinons, the corresponding Gutziller ground-state is
then unchanged after the flux attachment
$\vert \Psi_{\Phi }\rangle' = \vert \Psi_{0}\rangle $.  If we now
revert back to the original gauge, since $U_{\mu}f\dg_{\bk \sigma
}U_{\mu}^{-1}= f\dg_{\bk + \vA^{\sigma },\sigma }$, it follows that
\begin{equation}\label{eqlargeNGS}
\vert  \Psi_{\Phi }\rangle  = U_{\mu}\vert \Psi_{\Phi }\rangle'=
P_{G} \prod_{\bk \in {\rm FS}, \sigma }
f\dg_{\bk+ \vA^{\si}, \sigma }\vert  0 \rangle.
\end{equation}
corresponding to a Fermi sea in which the spinon momenta are shifted
by $\vA^{\sigma }= 2\pi/L_{x}\delta^{\sigma \mu}\hat x$, i.e
\begin{equation}\label{}
\Delta  P_{x} = \frac{2\pi}{L_{x}}
V\frac{V_{\rm FS}^\mu}{(2\pi )^{D}}
\end{equation}

By comparing this result with \eqref{eqdelPsl},
we then obtain
\begin{equation}
    \frac{V_{\rm FS}^\mu}{(2\pi)^D} =(m_\mu + q), \qquad
    \mathrm{mod}\ 1. \label{eqspinon}
\end{equation}
We emphasize that this result remains valid at arbitrary $N$ as long as the ground state is smoothly connected to the U(1) spin liquid state in \eqref{eqlargeNGS}.

\section{Discussion}\label{secDisc}

It is interesting to consider the implications of our results for the
FL$^*$ phase of the Kondo lattice model, in which decoupled spin
liquid and conduction electrons co-exist in a state of unbroken symmetry. 
Earlier work on $S=1/2$ Kondo
systems~\cite{senthil2003,paramekanti2004a} has interpreted this phase
as a Z$_2$ spin liquid coexisting with a Fermi liquid. Flux insertion
then drives a transition between two topologically degenerate
ground-states characterized by the presence or absence of vizon states
that carry Z$_2$ flux. But is the the FL$^*$ phase necessarily
topologically ordered?

Our result on the Kondo Heisenberg model suggests an alternate
interpretation of the FL$^{*}$ phase as the co-existence of a U(1) spin liquid 
with an electronic Fermi liquid. There are in principle, two phases:
\begin{itemize}
\item the heavy Fermi liquid, a Higgs phase in which the U(1)
gauge field of the spinons is locked to the electromagnetic U(1)
fields of the conduction electrons, giving rise to a single unified
Fermi surface of heavy electrons. 

\item the FL$^{*}$ in which the U(1) gauge fields of the conduction electrons and
spinons are decoupled, so that one is neutral, the other charged

\end{itemize}
Oshikawa's theorem, extended to the Kondo Heisenberg
model makes no judgement on which phase one is in, simply predicting
that the combined volume of the Fermi surfaces
\begin{equation}\label{}
\frac{V_{{\rm FS}}^\mu}{(2\pi)^{D}}
 = \frac{V^{\mu,S}_{\rm FS}}{(2\pi)^{D}} +
 \frac{V^{\mu,e}_{\rm FS}}{(2\pi)^{D}}
  = N_{s} (m_{\mu}+q) + \nu^{\mu}
\end{equation}
If the spin liquid decouples from the electronic fluid, 
then assuming that the U(1) spin liquid is isomorphic to that of the pure
Heisenberg model in Section~\ref{secSpinLiq}, the 
volume of the spinon Fermi surface is given by $\frac{V^{\mu,S}_{\rm
FS}}{(2\pi)^{D}}= N_{s} (m_{\mu}+q)$. 
In this case, the remaining electronic fluid has a Fermi surface volume
\begin{equation}\label{}
 \frac{V^{\mu,e}_{\rm FS}}{(2\pi)^{D}}= \nu_{\mu}.
\end{equation}
From this perspective, 
the FL$^*$ is understood simply as two decoupled fluids, both
of which respond to the flux attachment. One of the interesting
aspects of this line of reasoning, is that it goes against a commonly
held view-point that fractionalization in higher dimensional systems is
intimately associated with a topological ground-state.  It suggests
instead that fractionalization does not require such inevitable linkage,
and it opens the way for an interpretation of the Kondo effect as a
non-topological fractionalization of local moments.

Such U(1) spin liquids are expected from large $N$ treatments~\cite{coleman1986,affleck1988,vojta1999,hermele2004,coleman2005} and found in variational studies of Heisenberg-related models~\cite{motrunich2005}. Tantalizing evidence of the anomalous signatures in thermal conductivity~\cite{lee2005}, spin susceptibility~\cite{motrunich2005} and anomalous quantum oscillations expected of such spin liquids have been observed in experiments~\cite{li2014,tan2015,hartstein2018}.

One of the unsolved questions, is whether Oshikawa's approach can be
extended to other models?
Central to the current derivation of the Luttinger sum rule is the
identification of a U(1) gauge symmetry associated with \emph{each}
of the $N$ spin components, and the presence of  translational
symmetry. There are two models that fail these requirements:
\begin{itemize}
\item the
Kondo impurity model, where fractionalization, and the large $N$ limit
tell us that the scattering phase shift is given by $\delta = \pi Q/N$~\cite{coleman2016}.

\item the family of symplectic SP$(2N)$ symmetric Kondo lattices,
important for extending the notion of pairing to the large $N$ limit~\cite{read1991,flint2008}.

\end{itemize}
At first sight, the absence of a conserved momentum would seem to
preclude using flux attachment on the impurity Kondo model,
however however, by representing the impurities as
left moving particles in a fluid of right-moving electrons, as in
Bethe-Ansatz solutions of this problem~\cite{coleman1986},
it may be possible to restore translational invariance required for
flux attachment.

Likewise, the absence of a large number of U(1) subgroups in SP$(2N)$
appears to sabotage the application of Oshikawa's theorem to this
case. However, here too, there may be a way out, for the total number
of ``up'' electrons and spinons is still a conserved U(1) invariant,
so that if we attach a
flux to all the up electrons and spinons, a Fermi surface sum rule may
still be possible.
These topics can be considered in  future work.

\section*{Acknowledgments} \label{secAcknowledge}

This work was supported by NSF grant DMR-1830707 (PC, TH).

\bibliography{GeneralHeavyFermion}

\appendix
\section{Stability of the U(1) spin liquid in the large $N$ limit}\label{appMFT}

\fg{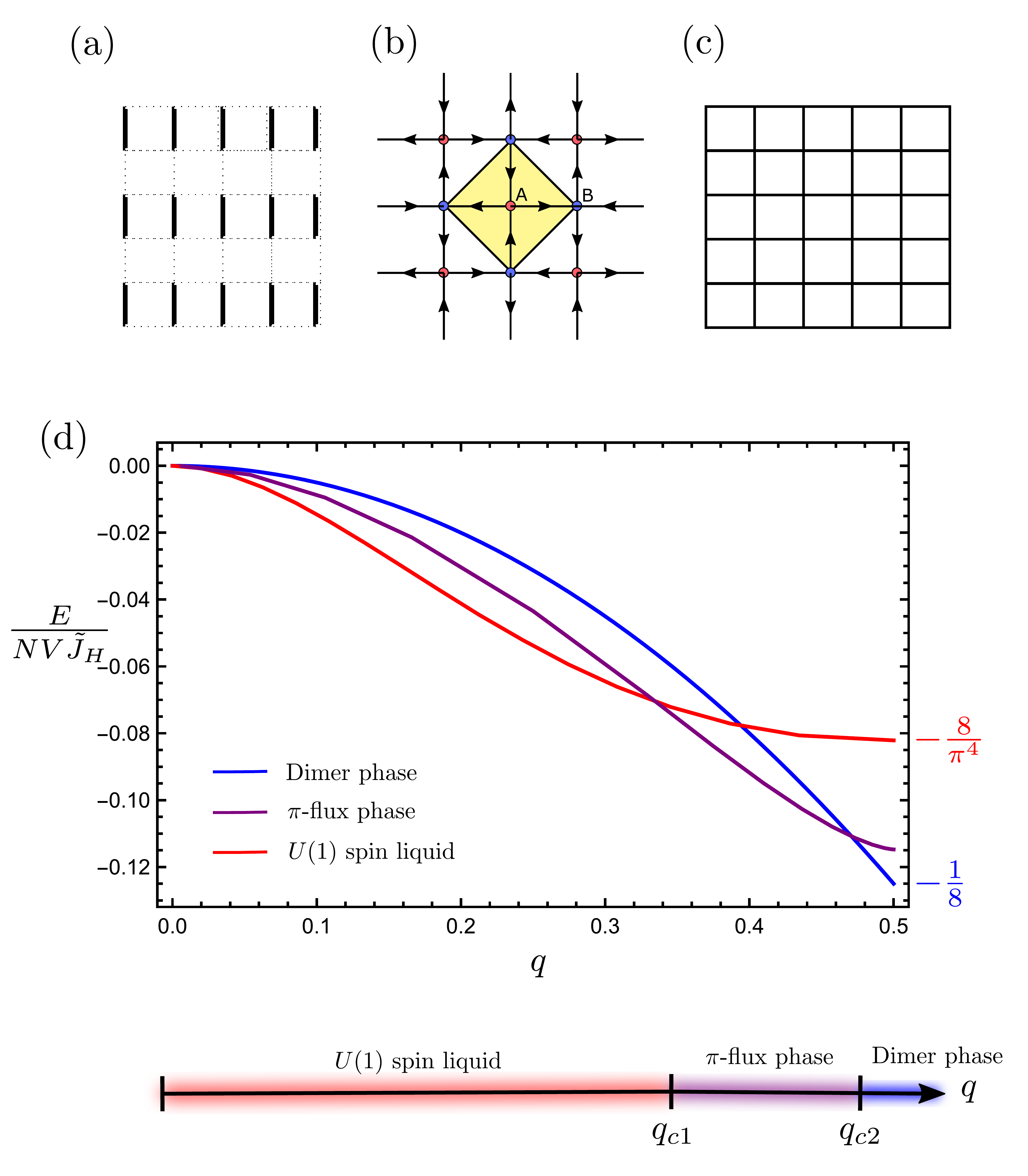}{figAppendix}{
(a) $\pi$-flux phase: the phase of the bond order parameter $\chi$ is positive along the direction of the arrows. The unit cell (yellow) is expanded to include two inequivalent sites $A$ and $B$, corresponding to a reduced Brillouin zone.
(b) Peierls phase in which the spin on each site forms a dimer with its nearest-neighbor and decouples from the lattice.
(c) Comparison of ground state energies for the spin liquid (red), flux phase (purple) and dimer phase (blue). The dashed curve is an analytical approximation for the spin liquid ground state energy valid at small $q$. For a range of filling $q<q_{c1}\sim 0.3$ the uniform U(1) spin liquid is stable with respect to the flux and dimer phases.
}

\begin{table*}
\caption{\label{tabMFResults} Summary of large $N$ mean-field results}
\begin{ruledtabular}
\begin{tabular}{l|l|l|l}
State& Fermionic excitations & Gap equation & Ground state energy \\
\hline
Dimer & $\epsilon_\pm = \lambda \mp \chi$ & $\chi = \Jt q$ & $\frac{2E}{\Jt N L^2}= -q^2$ \\
$\pi$-Flux & 
$\epsilon_{\bk \pm} = \lambda \pm 2\chi \sqrt{\cos^2 k_x +\cos^2 k_y}$ & 
$\frac{2|\chi|}{\tilde{J}_H}=\int \frac{d^2\bk}{(2\pi)^2} n_{\bk-} \sqrt{\cos^2 k_x +\cos^2 k_y}$ & 
$\frac{2E}{\Jt NL^2}=- \left( \int \frac{d^2\bk}{(2\pi)^2} n_{\bk-} \sqrt{\cos^2 k_x +\cos^2 k_y} \right)^2$  \\
U(1) SL & 
$\epsilon_\bk=\lambda - 2 \chi (\cos k_x +\cos k_y)$ & 
$\frac{2|\chi|}{\tilde{J}_H}=\int \frac{d^2\bk}{(2\pi)^2} (\cos k_x + \cos k_y)\, n_\bk^f$ & 
$\frac{2E}{\Jt NL^2}= -\left( \int \frac{d^2\bk}{(2\pi)^2} (\cos k_x + \cos k_y)\, n_\bk^f \right)^2$
\end{tabular}
\end{ruledtabular}
\end{table*}

The nearest-neighbor Heisenberg model is described by the path integral
\begin{align}
    \mathcal{Z}&=\int \mathcal{D}[f^\dagger,f^\phdag,\lambda] \exp\bigg[-\int_0^\beta d\tau\, \mathcal{L}(\tau) \bigg] \no \\
	\mathcal{L}&=\sum_{\br} \left(f^\dagger_{\br\si} (\partial_\tau +\lambda_\br ) f^\phdag_{\br\si} - \lambda_\br Q \right) + H_H,
\end{align}
with a summation convention over spin indices $\sigma= (1,N) $. The 
Heisenberg Hamiltonian $H_H$ represented in terms of the
Abrikosov fermi fields  $f^\dagger,f$ as
\begin{align}
	H_H = -\frac{\tilde{J}_H}{N} \sum_{\langle \br\br'\rangle}
	(f_{\br\si}^\dagger f_{\br'\si}^\phdag) ( f_{\br'\si'}^\dagger f_{\br'\si'}^\phdag),
\end{align}
while the constraint on the local fermion number is implemented by an
integral over the chemical potential
$\lambda_\br$~\cite{coleman2016}. Decoupling the four-fermion term by
a Hubbard-Stratonovich transformation to the resonating valence bond
fields $\chi_{\br\br'}$ and approximating the integral by the saddle
point action leads to the mean-field Hamiltonian
\begin{align}\label{eqMFgen}
    H_{MF}&=-\sum_{\langle \br\br'\rangle} \chi_{\br\br'} f_{\br\si}^\dagger f_{\br'\si}^\phdag +\frac{N}{\tilde{J}_H} \sum_{\langle \br\br' \rangle} |\chi_{\br\br'}|^2 \no \\
    &\qquad+\sum_{\br} \lambda_\br (f_{\br\si}^\dagger f_{\br\si}^\phdag - Q),
\end{align}
which becomes exact in the limit of large $N$. We compare the energies of the spin-liquid (SL), dimer, and $\pi$-flux phases of this Hamiltonian on a square lattice in two dimensions with linear dimension $L$ in units of the lattice constant. Each of these phases has $\lambda_\br=\lambda$.

For the dimer or Peierls phase~\cite{majumdar1969}, $\chi_{\br\br'}=0$
on all but one of the nearest-neighbor bonds to each site, as shown in
Fig.~\ref{figAppendix}(a). For the $\pi$-flux
phase~\cite{affleck1988}, $\chi_{\br\br'}=|\chi| e^{i\pi/4}$ if
$\br\rightarrow\br'$ is oriented along the arrows in
Fig.~\ref{figAppendix}(b). For the uniform U(1) spin liquid,
$\chi_{\br\br'}=\chi \in \mathbb{R}$ for all bonds
(Fig.~\ref{figAppendix}(c)). Table~\ref{tabMFResults} summarizes the
results of the large $N$ mean-field analysis for these states when
$q=Q/N\le1/2$ and the ground state energies are compared in
Fig.~\ref{figAppendix}(d). Near half-filling, the Peierls phase has
the lowest energy. However, for low filling upto $q\sim0.3$, the
lowest energy state is the uniform U(1) spin liquid. For intermediate
filling $0.3<q<0.48$, the flux phase is most stable.

When $q\ll 1$, the dispersion of the filled states is approximately quadratic and we obtain the following analytical expressions for the ground state energy
\begin{equation}\label{}
\frac{E}{NV\Jt} = \left\{
\begin{array}{lr}
 -\left[ \frac{1-(1-2\pi q)^{3/2}}{3\pi} \right]^2  \simeq -q^2, &\text{$\pi$-Flux} \cr\cr
 -q^{2}/2 , &\text{Dimer}  \cr\cr
-2q^2\left( 1- \frac{\pi q}{2} \right)^2 \simeq -2q^2, &\text{SL}
\end{array}
 \right.
\end{equation}
so for small $q$, the uniform spin liquid is the most energetically
favorable state. We note that while the dimer phase is stable only
near $q=1/2$, similar phases may be present and favorable at other
rational fillings. For instance, $r$-site ring polymer
states have energy $E/(N V \Jt) =-q^2$ at $q=1/r$. At $q=1/4$, the
4-site plaquette states have lower energy than the uniform U(1) spin
liquid. As $q$ becomes smaller, the likely ground state involves
larger and larger decoupled clusters with vanishing energy differences
$\Delta E$ from the U(1) spin liquid. Above temperatures of the order
of $\Delta E$, the system behaves like a spin liquid. Additionally, on
finite-sized systems, incommensuration between the cluster size and
the system size may frustrate the valence bond crystal and favor the
spin liquid.

\section{Flux insertion in the large-$N$ limit of Heisenberg model}\label{appFlux}

In this section, we explicitly demonstrate the flux insertion and
concomittant change in momentum in the Heisenberg model on a square
lattice in the limit of large $N$, in terms of the mean-field
Hamiltonian~\eqref{eqMFgen}. We discuss the response of the global
U(1) gauge corresponding to the phase of the bond order parameters
to the insertion of the flux, and explicitly show that the volume of
the spinon Fermi surface is given by \eqref{voila2}. As opposed to the
main text, we consider a flux that couples to $P$ of the $N$ spin
degrees of freedom, so that the effect of the flux threading on the
relative change in the ground state energy, for instance, is
non-vanishing in the large-$N$ limit.

The Heisenberg model in presence of such a flux is given by
\begin{align}
	H_H [\Phi_{\{\mu\}}]&=-\frac{\Jt}{N}\sum_{\langle\br\br'\rangle} (f_{\br\si}^\dagger f_{\br'\si}^\phdag) (f_{\br'\si'}^\dagger f_{\br'\si'}^\phdag\no) \\
		&\qquad \qquad \qquad \times e^{-i (\vA^\si -\vA^{\si'}) \cdot (\br-\br')} 
\end{align}
where $\vA^\si = (2\pi/L_x) \hat{x} \sum_{\{\mu \}} \delta_{\si \mu}$ with
the sum over $P$
spin-channels
to which the flux is coupled, where $\{\mu \}= \{\mu_{1}\dots ,\mu_{P}\}$,. In the large $N$ limit, this is exactly captured by the mean-field Hamiltonian
\begin{align}
	H_{MF}[\Phi_{\{\mu\}}]&=- \sum_{\langle \br \br'\rangle } \chi_{\br\br'} e^{-i \vA^{\si} \cdot (\br-\br')} f_{\br \si}^{\dagger} f_{\br' \si} \no  \\
	&+\frac{N}{\tilde{J}_H} \sum_{\langle \br\br' \rangle} |\chi_{\br\br'}|^2 +\sum_{\br} \lambda_\br (f_{\br\si}^\dagger f_{\br\si}^\phdag - Q)
\end{align}
The saddle point condition for a uniform order parameter leads to the self-consistency equation
\begin{align}
	\chi_{\br\br'}=|\chi| e^{-i\cA \cdot (\br-\br')} = \frac{\Jt}{2NV} \sum_{\langle \br\br'\rangle} \langle f_{\br'\si}^\dagger f_{\br\si}^\phdag \rangle e^{i\vA^\si \cdot (\br-\br')}
\end{align}
where $\cA$ is the global (spin-independent) U(1) gauge potential, $V$ is the volume of the system (with the unit cell volume set to unity). With this, the mean-field Hamiltonian is diagonal in momentum space
\begin{align}
	H_{MF}[\Phi_{\{\mu\}}]&=\sum_{\bk} \epsilon_{\bk+\cA+\vA^\si} f_{\bk\si}^\dagger f_{\bk\si}^\phdag 
	&+\frac{2NV|\chi|^2}{\Jt} -\lambda Q V
\end{align}
where $\epsilon_\bk=-2\chi (\cos k_x +\cos k_y) +\lambda$ is the
dispersion of the $f$-fermions. Recall that the ground state is the
same as before the flux insertion, since momentum is conserved
throughout the process. The final state $|\psi_\Phi \rangle' =
\Pi_{\bk\in {\rm FS},\si} f_{\bk\si}^\dagger |0\rangle$ and the
canonical x-momentum $P_x^0 = \sum_{\bk\in {\rm FS},\si} \bk =0$ as
before the flux insertion.  The gauge transformation that removes the
flux is now given by 
\begin{eqnarray}\label{l}
U^{s}_{\{\mu  \}}
&=& \exp \left[ \frac{2\pi i }{L_{x}}\sum_{\br, \{\mu \}} x_{\br }
 (\Lambda_{\br}^{\mu}+q)\right]
.
\end{eqnarray}
To transform back to the original gauge we
note that 
\begin{eqnarray}\label{l}
U^s_{\{\mu\}} f_{\bk\si}^\dagger U^{s\dagger }_{\{\mu\}} &=&
\frac{1}{\sqrt{V}} \sum_\br U^s_{\{\mu\}} f_{\br\si}^\dagger
U^{s\dagger}_{\{\mu\}} e^{i \bk \cdot \br}\cr
& =& \frac{1}{\sqrt{V}} \sum_\br
f_{\br\si}^\dagger e^{i (\bk+\vA^\si) \cdot \br} =
f_{\bk+\vA^{\si} \si}^\dagger. \no \\
\end{eqnarray}
We transform the ground state to this gauge
$|\psi_\Phi \rangle \equiv U^s_{\{\mu\}} |\psi_\Phi \rangle' =
\Pi_{\bk\in {\rm FS},\si} f_{\bk+\vA^\si,\si}^\dagger |0\rangle$ and
evaluate the physical momentum
\begin{align}\label{eqthat}
	P_x= \sum_{\bk\in {\rm FS},\si} (\bk + \vA^\si) = \sum_{\bk\in
	{\rm FS},\si} \vA^\si = P V \frac{V_{FS}}{(2\pi)^2} \frac{2\pi}{L_x}
\end{align}
when $P$ out of $N$ spin-components are coupled to the flux. In this case, the change in momentum on flux insertion can be independently computed following the arguments in the main text (\eqref{l1}-\eqref{eqdelPsl}) to yield
\begin{align}
	\Delta P_x =  V \left(\frac{2\pi}{L_x} \right) Pq
\end{align}
when the ground state is unpolarized. Comparing with \eqref{eqthat}, we find the volume of the Fermi surface to be $V_{FS}=(2\pi)^2 q$ consistent with \eqref{eqspinon}.

As the flux is inserted, the global U(1) gauge potential $\cA$ adjusts
in response to preserve a zero total spinon current. Symmetry dictates that $\cA \parallel \hat{x}$ and the new self-consistent value of $\cA$ is determined by the saddle point condition $\partial_\cA E=0$ leading to 
\begin{align}
	&\sum_{\bk\si} (\partial_\cA \epsilon_{\bk+\cA+\vA^\si}) n_\bk^f \no \\
	&= -2|\chi|\frac{\partial}{\partial_\cA} \sum_{\bk\si} \cos (\cA+\vA^\si) \cos k_x n_\bk^f \no  \\
	&= 2|\chi| \left(\sum_\bk \cos k_x n_\bk^f \right) \left(\sum_\si \sin (\cA +\vA^\si)\right) =0 \\
	&\Rightarrow P \sin(\cA- \frac{2\pi}{L_x})+ (N-P)\sin \cA =0
\end{align}
Since $L_x \gg 1$, we find that the saddle point value of $\cA$ is 
\begin{align}
	\cA=-\frac{2\pi}{L_x} \rho ,
\end{align}
where $\rho =\frac{P}{N}$
The global U(1) gauge potential adjusts to oppose the inserted flux and is proportional to the fraction of spin-components coupled to the flux. In fact, this keeps the net charge current fixed at 0, as expected for a response to a spin-twist. The flux imparts momentum to the $\mu$-fermions and elicits a diamagnetic response from all the fermions. This can be seen explicitly by calculating the ground state energy in presence of the flux
\begin{align}
	&E-\frac{2NV |\chi|^2}{\Jt} \no\\ 
	&= -2|\chi| \sum_{\bk\si} \left( \cos (k_x + \cA +A^\si_x) +\cos k_y \right) n_\bk^f \no \\
	&= -2|\chi| I_0 V \sum_{\si} \left( \cos (\cA +A^\si_x) + 1 \right)
\end{align}
where $I_0 \equiv (1/V) \sum_{\bk} \cos k_x n_\bk^f$. When $P$ of the $N$ spin-components couple to the flux,
\begin{align}
	\frac{E}{V} = -2|\chi|I_0 &\big( P \cos (\cA + A_x) \no \\
	&+ (N-P) \cos \cA - N \big) + \frac{2N|\chi|^2}{\Jt}
\end{align}
With $\cA \approx -\rho A_x$, the saddle point condition
$\partial_{|\chi|} E=0$ yields $|\chi| = (1/2) I_0 \Jt (\rho
\cos((1-\rho)A_x) +(1-\rho) \cos (\rho A_x) +1)$.
As a result, the ground state energy is 
\begin{align}
	\frac{E}{N V} &=-2 |\chi|^2/\Jt \no \\
	&= -\frac{1}{2} \Jt I_0^2 \big(\rho (\cos ((1-\rho)A_x) \no \\
	&\qquad\quad+ (1-\rho) \cos (\rho A_x) +1\big)^2 \no \\
	&\approx -2 \Jt I_0^2 \left[ 1- \frac{1}{8}\left(\rho(1-\rho)^2 +(1-\rho)\rho^2 \right) A_x^2  \right]^2 \no \\ 
	&= \frac{E_0}{N V} + \frac{1}{2} \rho \widetilde{D}_{\{\mu\}}  A_x^2 + O(A_x^4)
\end{align} 
where $E_0 = -2 N V \Jt I_0^2$ is the energy in absence of the flux
and $\widetilde{D}_{\{\mu\}} = 2 \Jt I_0^2  (1-\rho)$. The
quantity $j_{x} = - (1/V) \partial E/\partial A_{x}= - P\tilde{D}_{\{\mu \}}
A_{x}$  can be interpreted as the
diamagnetic spin current response of the P-spin channels 
to the flux insertion, while $\widetilde{D}_{\{\mu\}}$ 
is their spin stiffness. 

\end{document}